\documentclass[a4paper,11pt]{article}
\pdfoutput=1 

\usepackage{jcappub} 

\usepackage[T1]{fontenc} 

\newcommand{\Mpl}{M_{\textrm{Pl}}}
\renewcommand{\(}{\left(}
\renewcommand{\)}{\right)}

\newcommand{\nn}{\nonumber}
\def\al{\alpha}
\def\bet{\beta}

\def\Om{\Omega}
\def\sig{\sigma}

\def\lam{\lambda}

\def\S{\mathcal{S}}

\def\doi{http://doi.org}

 \def\e{\mathrm{e}}
\def\r{\mathrm{r}}
\def\g{\mathrm{g}}

\def\m{\mathrm{m}}

\def\d{\mathrm{d}}

\allowdisplaybreaks

\usepackage{orcidlink}

\title{\boldmath General parametrization for energy density of quintessence field}

\author[1]{Shiriny Akthar\note{Current Institute: Department of Astronomy, Astrophysics \& Space Engineering, Indian Institute of Technology Indore, Indore 453552, India.}}
\author{and Md. Wali Hossain\orcidlink{0000-0001-6969-8716}}
\affiliation{Department of Physics, Jamia Millia Islamia,\\ New Delhi, 110025, India}

\emailAdd{shirinyakthar@gmail.com}
\emailAdd{mhossain@jmi.ac.in}

\abstract{We present a general parametrization for energy density of a quintessence field, a minimally coupled canonical scalar field which rolls down slowly during the late time. This parametrization can mimic all classes of quintessence dynamics, namely scaling-freezing, tracker and thawing dynamics for any redshift. For thawing dynamics the parametrization needs two free parameters while for scaling-freezing and tracker dynamics it needs at least four free parameters. More parameters make the model less interesting from the observational data analysis point of view but as we expect more precise data in future it may be possible to constrain the models with multiple free parameters which can tell about the dynamics more precisely. One of the main advantage of this parametrization is that it reduces the computational time to significant amount while mimicking the actual scalar field dynamics for all redshifts which may not be possible with other existing parametrizations. We compare the parametrization with two and four parameters with the standard $\Lambda$CDM model, $w$CDM and Chevallier-Polarski-Linder (CPL) parametrizations using cosmological observational data from Planck 2018 (distance priors), DESI $2024$ DR1, PantheonPlus, Hubble parameter measurements and the redshift space distortion. We find that the observational data prefers standard $\Lambda$CDM model over other models. If we allow phantom region then it is more preferred by the data compared to non-phantom thawing quintessence. Our analysis does not show any preference of the dynamical dark energy over a cosmological constant except for the CPL parametrization.}


\arxivnumber{arXiv:2411.15892}

\begin{document}
\maketitle
\flushbottom

\section{Introduction}
\label{sec:intro}
After Planck 2013 results \cite{Planck:2013pxb} the almost cosmological constant (CC) appeared not only as the most favoured but the sufficient candidate as the dark energy with constant equation of state (EoS) to explain the late time universe. So, the relevance of dynamical dark energy (DDE) \cite{Copeland:2006wr,Bahamonde:2017ize} was almost became irrelevant. it is very recently, after the local measurement of the present value of the Hubble constant ($H_0$) by the SH0ES team \cite{Riess:2021jrx}, which observed $\sim 5\sig$ tension between the local measurement of $H_0$ and the constraint coming from cosmic microwave background (CMB) observations assuming the standard $\Lambda$CDM model \cite{Planck:2018vyg}, the DDE becomes important. Apart from the Hubble tension \cite{Kamionkowski:2022pkx,Freedman:2023jcz,Bernal:2016gxb,Knox:2019rjx} the standard $\Lambda$CDM model is also in tension in the measurements of growth rate known as the $S_8=\sig_8 \sqrt{\Om_{\m0}/0.3}$ tension \cite{Perivolaropoulos:2021jda,Kilo-DegreeSurvey:2023gfr}, where $\sig_8$ is the standard deviation of
matter density fluctuations at present for linear perturbation in spheres of radius $8h^{-1}{\rm Mpc}$ and $\Om_{\m0}$ is the present value of matter density parameter. On top of these tensions the measurements of baryon acoustic oscillation (BAO) by the DESI experiment in 2024, combined with other observational data, shows a preference on DDE over CC \cite{DESI:2024mwx,DESI:2024aqx,DESI:2024kob,Giare:2024smz,Berghaus:2024kra,Qu:2024lpx,Wang:2024dka,Giare:2024gpk,Shlivko:2024llw,Ye:2024ywg,Ramadan:2024kmn,Jiang:2024xnu,Payeur:2024dnq,Pourojaghi:2024tmw,Wolf:2023uno,Wolf:2024eph,Wolf:2024stt,Chan-GyungPark:2024spk,Chan-GyungPark:2024brx,Colgain:2024xqj}. So, it's a high time to study DDE.

One of the simplest candidate of DDE is the minimally coupled canonical scalar field, $\phi$. If this scalar field rolls slowly during the late time we call it a quintessence field \cite{Ratra:1987rm,Wetterich:1987fk,Wetterich:1987fm} which can explain the late time acceleration \cite{Copeland:2006wr}. In a cosmological background, the scalar field EoS $w_\phi$ varies between $-1$ to 1. $w_\phi=-1$ signifies potential energy domination while $w_\phi=1$ signifies kinetic energy domination. The energy density of the scalar field $\rho_\phi$ can be represented as $\rho_\phi\sim \e^{-3\int (1+w_\phi)\d \ln a}$. $w_\phi=-1$ gives constant $\rho_\phi$ and it varies as $a^{-6}$ for $w_\phi=1$. Thus, we can expect wide variation in the scalar field dynamics in a cosmological background. In fact, we can classify the dynamics in three classes, namely, scaling-freezing \cite{Copeland:1997et,Barreiro:1999zs,Sahni:1999qe}, tracker \cite{Steinhardt:1999nw,Zlatev:1998tr} and thawing \cite{Scherrer:2007pu} models. Because of the large Hubble damping coming from the background energy density in the scaling-freezing and tracker dynamics the scalar field is frozen in the past. During the frozen period $\rho_\phi$ becomes almost constant while the background energy density decreases. When $\rho_\phi$ becomes comparable to the background energy density, the Hubble friction weakens, allowing the scalar field to roll down its potential once again. Depending upon the nature of the potential the scalar field energy density can either scale the background energy density (for scaling-freezing) or decay a little slower than the background energy density (for tracker). For tracker, since the decay of scalar field energy density is slower than the background the scalar field eventually takes over matter during the recent past which may give rise to late time acceleration with viable cosmology\cite{Hossain:2023lxs}. For scaling-freezing the scalar field scales the background after the frozen period and then takes over matter in the recent past \cite{Barreiro:1999zs} which can be achieved either by having a sufficiently shallow region in the potential during the late time or by having a nonminimal coupling between the scalar field and matter in the Einstein frame. For both the dynamics the late time dynamics is an attractor solution. On the other hand the thawing dynamics is very sensitive to the initial conditions. In this dynamics the scalar field is frozen from the past behaving like a CC which starts to thaw from the recent past giving rise to deviation from CC \cite{Scherrer:2007pu}.

While scalar field gives interesting dynamics in the cosmological background it can be very expensive computationally while running simulations to analyse cosmological data. For this reason it is always helpful to work with parametrized form of EoS or energy density of the DDE. In this regard the  Chevallier-Polarski-Linder (CPL) \cite{Chevallier:2000qy,Linder:2002et} parametrization of EoS of DDE works very well for low redhsifts with only two parameters and this is the most widely used parametrization. Apart from the CPL parametrization many other parametrization have been studied in the literature {\it e.g.}, logarithmic (Efstathiou model) \cite{Efstathiou:1999tm}, Jassal-Bagla-Padmanabhan \cite{Jassal:2005qc}, Barboza-Alcaniz \cite{Barboza:2008rh} parametrization. Generally these parametrizations can mimic DDE at very low redshifts. Now, introduction of scalar field gives actual model of DDE which can have some specific nature in dynamics and may not be mimicked by any arbitrary parametrization specially at higher redshifts. So, in this paper, we present a general parametrization of the energy density of quintessence field which not only can mimic the cosmological dynamics of a quintessence field for a particular potential but also can be considered as a model-agnostic framework for studying dynamical dark energy. This enables us to explore a broad class of quintessence models with or without assuming specific forms of the scalar field potential. 

The main challenge for this general parametrization is to reduce the number of free parameters. In this regard, we have deduced some relations to reduce the number of free parameters. The thawing dynamics can be represented with two free parameters while the scaling-freezing and tracker dynamics can be represented with at least four parameters. Now, having more parameters makes the scenario less interesting from the data analysis point of view but these parameters are needed if we are interested specifically in the scalar field dynamics. We expect that in near future we will have more precise data which can constrain these parameters and we will be able to tell about the scalar field dynamics more precisely. Also, working with the scalar fields can make the computation, for data analysis, very expensive in time. This general parametrization reduces the computational time to significant amount while preserving the actual scalar field dynamics for a particular potential for any redshift. This is one of the main advantage of this parametrization.

The paper is organised as follows. In Sec.~\ref{sec:background} we introduce the background cosmological equations including scalar field equation of motion. In Sec.~\ref{sec:para} we introduce the general parametrization of the energy density of a quintessence field. The scaling-freezing dynamics has been studied in Sec.~\ref{sec:scaling-freezing}. The tracker dynamics has been studied in Sec.~\ref{sec:tracker} while in Sec.~\ref{sec:thawing}. The study of observational constraints has been done in Sec.~\ref{sec:obs}. We summarise and conclude our results in Sec.~\ref{sec:conc}.

\section{Background equations}
\label{sec:background}
We consider a minimally coupled canonical scalar field with the following action
\begin{align}
\S=\int \d^4x\sqrt{-\g}\Bigl [\frac{\Mpl^2}{2} R-\frac{1}{2}\partial_\mu\phi\partial^\mu\phi - V(\phi) \Bigr]+ \S_\m+\S_\r \, ,
\label{eq:action}
\end{align}
where $\Mpl=1/\sqrt{8\pi G}$ is the reduced Planck mass and $V(\phi)$ is the potential of the field. $\S_\r$ and $\S_\m$ are the actions for radiation and matter respectively.

Varying the action (\ref{eq:action}) with respect to (w.r.t.) the metric $g_{\mu\nu}$ gives the Einstein's field equation
\begin{align}
\Mpl^2 G_{\mu\nu}= T_{(\m)\mu\nu}+T_{(\r)\mu\nu}+T_{(\phi)\mu\nu} \, ,
\label{eq:ee}
\end{align}
where
\begin{align}
T_{(\phi)\mu\nu}=& \phi_{;\mu}\phi_{;\nu}-\frac{1}{2}\g_{\mu\nu}(\nabla\phi)^2 -\g_{\mu\nu}V(\phi) \, .
\label{eq:emt_phi}
\end{align}
The equation of motion of the scalar field can be calculated by varying the action (\ref{eq:action}) w.r.t. the scalar field $\phi$ and it is given by  
\begin{align}
& \Box \phi-V_\phi(\phi)=0 \, ,
\label{eq:eom_phi}
\end{align}
where subscript $\phi$ denotes the derivative wrt $\phi$.

In flat  Friedmann--Lema\^itre--Robertson--Walker (FLRW) metric, given by
\begin{align}
\label{metric0}
 ds^2 = -dt^2 +a(t)^2\delta_{ij} dx^idx^j ~,
\end{align}
where $a(t)$ is the scale factor, the Friedman equations are given by
\begin{eqnarray}
 3H^2\Mpl^2 &=& \rho_\m+\rho_\r+\frac{1}{2}\dot\phi^2+V(\phi) \,  
 \label{eq:Friedmann1}\\
 \(2\dot H+3H^2\)\Mpl^2&=&-\frac{1}{3}\rho_\r-\frac{1}{2}\dot\phi^2+V(\phi) \, .
 \label{eq:Friedmann2}
\end{eqnarray}
The equation of motion of the scalar field is given by
\begin{equation}
 \label{scalareom1}
 \ddot\phi+3H\dot\phi+\frac{\d V}{\d\phi}=0 \, .
 \end{equation}

Effective equation of state (EoS) and the EoS of the scalar field are given by
\begin{eqnarray}
w_{\text{eff}}&=& -\(1+\frac{2}{3}\frac{\dot H}{H^2}\)  \; , 
\label{eq:weff}\\
w_{\phi} &=& \frac{\frac{1}{2}\dot\phi^2-V(\phi)}{\frac{1}{2}\dot\phi^2+V(\phi)}  \; .
\label{eq:wphi}
\end{eqnarray}

We define the following two functions 
\begin{eqnarray}
    \lambda &=& -\Mpl\frac{V'(\phi)}{V} \; , 
    \label{eq:lam}\\
    \Gamma &=& \frac{V''(\phi)V(\phi)}{V'(\phi)^2} \; .
    \label{eq:gam}
\end{eqnarray}
While the function $\lam$ signifies the slope of the potential the function $\Gamma$ represents the nature of the potential, {\it e.g.}, $\Gamma=1$ for an exponential potential of constant slope $\lam$. The nature of these functions determines the scalar field dynamics \cite{Bahamonde:2017ize,Hossain:2023lxs}.

\section{Scalar field dynamics and the parametrization}
\label{sec:para}

\begin{figure}[ht]
\centering
\includegraphics[scale=.93]{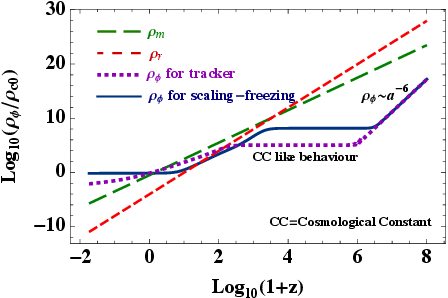} ~~~
\includegraphics[scale=.94]{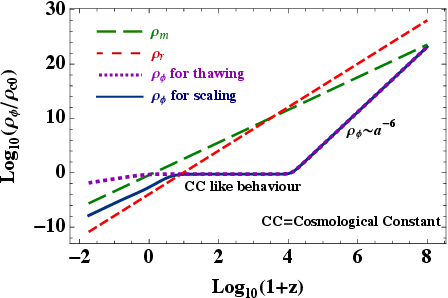}  \vskip10pt
\includegraphics[scale=.93]{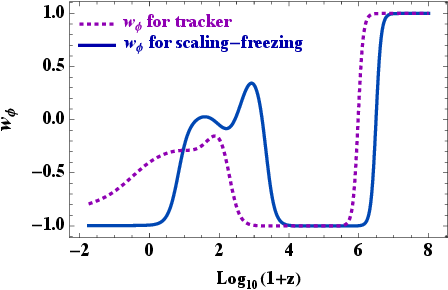} ~~~
\includegraphics[scale=.94]{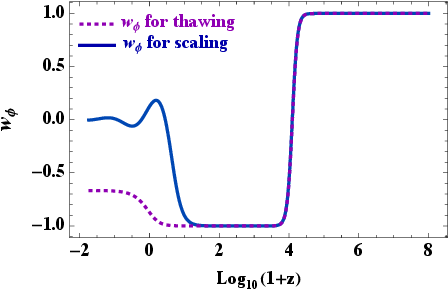} 
\caption{Different scalar field dynamics has been shown. Left figures show the tracker and scaling-freezing dynamics while the right figures show the thawing and scaling dynamics. Long dashed green line represents matter energy density while the short dashed red line represents radiation energy density. In left figures purple dotted lines represent tracker dynamics while blue solid lines represent scaling-freezing dynamics. In the right figures purple dotted lines represent thawing dynamics while blue solid lines represent scaling dynamics.} 
\label{fig:dynamics}
\end{figure}

Scalar field dynamics, represented in Fig.~\ref{fig:dynamics}, depends on the potential $V(\phi)$. More specifically it depends on the nature of the functions $\lam$ and $\Gamma$. Considering different kinds of scalar field dynamics we can classify them in three categories, (i) scaling-freezing \cite{Copeland:1997et,Copeland:2006wr}, (ii) tracker \cite{Zlatev:1998tr,Steinhardt:1999nw} and (iii) thawing dynamics \cite{Scherrer:2007pu,Caldwell:2005tm}. (i) The scaling-freezing dynamics, shown in the left figure of Fig.~\ref{fig:dynamics} with blue solid line, can be achieved in potentials with large slope $\lam$ while $\Gamma$ should be equal or very close to 1 for some region of the potential with large slope where the scalar field energy density will scale the background energy density. In this dynamics the scalar field remains frozen in the past, just before reaching the scaling solution, due to large Hubble damping and then scale the background during the intermediate time before taking over matter giving rise to late time acceleration. In this dynamics $\lam$ should vary from a large value to a smaller value to achieve viable cosmology. If $\lam$ remains constant and large then for $\Gamma=1$ we have only scaling solutions, represented by the blue solid line in the right figure of Fig.~\ref{fig:dynamics}, where the scalar field will scale the background forever and can not take over it \cite{Copeland:1997et}. (ii) Like the scaling-freezing dynamics in tracker dynamics also we need large slope of the potential but $\Gamma>1$ \cite{Zlatev:1998tr} with a frozen stage of the scalar field in the past just before reaching the tracker behaviour, shown in the left figure of Fig.~\ref{fig:dynamics} with the purple dotted line. As $\Gamma$ goes away from 1 towards the larger value, for large slope $\lam$, the scalar field does not scale the background exactly but decays a little slower than background which results an eventual domination of scalar field during the late time which may give rise to late time acceleration. The advantage of  both scaling-freezing and tracking dynamics is that the late time cosmology is an attractor solution can be similar for an wide range of initial conditions. The disadvantage is that the requirements of specific nature of the functions $\lam$ and $\Gamma$ put a constraint on the nature of the potential and not all potential can give rise to these dynamics. (iii) In thawing dynamics, shown in the right figure of Fig.~\ref{fig:dynamics} with the purple dotted line, the scalar field remains frozen until it starts evolving slowly from the recent past giving rise to late time acceleration. In this dynamics the scalar field behaves like a CC for most of the time except very recently when it starts evolving and deviates from the CC nature. To have viable cosmology the scalar field can not deviate much from the CC at present which can be achieved by tuning the initial conditions. So, this dynamics is very initial condition dependent but can be achieved for any potential. As the scalar field remains frozen for most of the time and then evolves slowly the scalar field dynamics may not capture the actual interesting features of the potential in this dynamics at least until the present time.

\begin{figure}[t]
\centering
\includegraphics[scale=.96]{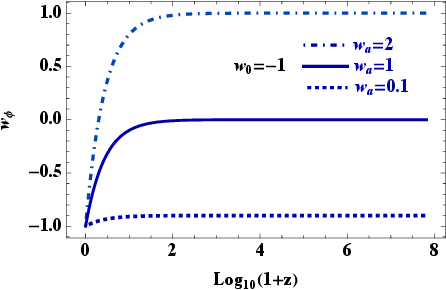} ~~~
\includegraphics[scale=.94]{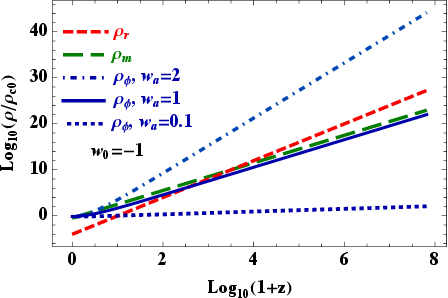} 
\caption{Evolution of EOS and energy density, in CPL parametrization, have been shown for $w_0=-1$ and $w_{\rm a}=0.1,1\; {\rm and}\; 2$.} 
\label{fig:cpl}
\end{figure}

From the above discussion we can understand that the energy density of the scalar field $\rho_\phi\sim a^{-n}$, where $n$ can vary from $6$ to $0$. This can also be understood from the fact that 
\begin{eqnarray}
    \rho_\phi \sim \e^{-3\int (1+w_\phi)d\ln a} \, ,
\end{eqnarray}
and from Eq.~\eqref{eq:wphi} we can see that 
\begin{eqnarray}
    w_\phi &\approx& 1 \, , ~~~ {\rm when}~~\frac{1}{2}\dot\phi^2\gg V(\phi)\, , ~~~~~ \Longrightarrow \rho_\phi \sim a^{-6} 
    \label{eq:rho_appr}\\
    w_\phi &\approx& -1\, ,  ~ {\rm when}~~\frac{1}{2}\dot\phi^2\ll V(\phi) \, , ~~~~~ \Longrightarrow \rho_\phi \sim a^0 \, .
\end{eqnarray}
This wide range of variation in $\rho_\phi$ or $w_\phi$ is difficult to capture while parametrizing the evolution of $\rho_\phi$ or $w_\phi$. {\it E.g.}, CPL parametrization is given by
\begin{eqnarray}
    w_{\phi,{\rm CPL}} = w_0+w_{\rm a} \frac{z}{1+z} \, ,
    \label{eq:cpl_wphi}
\end{eqnarray}
where $w_0$ and $w_{\rm a}$ are constants. Corresponding energy density is given by
\begin{eqnarray}
    \rho_{\phi,{\rm CPL}} = \rho_{\phi 0}(1+z)^{3(1+w_0+w_{\rm a})}\e^{-\frac{3w_{\rm a}z}{1+z}} \, .
    \label{eq:cpl_rho}
\end{eqnarray}
In the CPL parametrizaion, for $z>1$, we have $w_{\phi,{\rm CPL}}\approx w_0+w_{\rm a}$, {\it i.e.}, the scalar field EoS remains constant  at $w_0+w_{\rm a}$ and the scalar field energy density effectively behaves as $\rho_{\phi,{\rm CPL}}\approx \rho_{\phi 0}\e^{-3w_{\rm a}}(1+z)^{3(1+w_0+w_{\rm a})}$. Around $z=1$, $w_{\phi,{\rm CPL}}$ starts to deviate from the value $w_0+w_{\rm a}$ and becomes $w_0$ at $z=0$. As this transition happens around $z=1$ this transition is quite sharp which does not allow the scalar field energy density (Eq.~\eqref{eq:cpl_rho}) to have a particular nature for a sufficiently long period apart from the nature where $\rho_{\phi,{\rm CPL}}\approx \rho_{\phi 0}\e^{-3w_{\rm a}}(1+z)^{3(1+w_0+w_{\rm a})}$. Now, for $w_0=-1$ and $w_{\rm a}=2$ we have $w_{\phi,{\rm CPL}}\approx 1$ for $z>1$ and $\rho_{\phi,{\rm CPL}}\approx \rho_{\phi 0}(1+z)^6$ which follows Eq.~\eqref{eq:rho_appr}. But the problem is, as $\rho_{\phi,{\rm CPL}}\sim (1+z)^6$ until redshift around $z=1$, $\rho_{\phi,{\rm CPL}}$ always becomes larger than $\rho_\m$ and $\rho_\r$ which makes the scenario  non viable. We have depicted this in Fig.~\ref{fig:cpl}. Left figure of Fig.~\ref{fig:cpl} shows the evolution of the EoS of dark energy in CPL parametrization and the right figure shows the corresponding evolution of the energy density for different parameter values. If we compare the figures \ref{fig:dynamics} and \ref{fig:cpl} then it is very clear that the CPL parametrization can not reproduce all the dynamics of scalar field. In fact, the right figure of Fig.~\ref{fig:cpl} shows that $\rho_{\phi,{\rm CPL}}$ can be larger than the matter density from very low redshift for $w_{\rm a}>1$. For $w_0=-1$ and $w_{\rm a}=1$ we have $w_{\phi,{\rm CPL}}=0$, {\it i.e.}, in CPL parametrization, for $w_\phi>0$, we do not have a viable cosmology. Another important requirement is to have a frozen period during the high redshift for scaling-freezing and tracker models which can not be obtained in CPL parametrization. So, we need to formulate a general parametrization which can reproduce the required dynamics at every redshift to reproduce the full dynamics as shown in Fig.~\ref{fig:dynamics}.

Considering the above mentioned requirements to reproduce the scalar field dynamics we have to incorporate, in the parametrization, the changes in $\rho_\phi$ at different epochs. To do this we use the function 
\begin{eqnarray}
    \rho_i(z) = \frac{\rho_{0}}{1+\(\frac{1+z_{i}}{1+z}\)^{\al_i}} \, ,
\end{eqnarray}
with constant $\al_i$ and $z_i$. The above function behaves almost as a constant for $z>z_i$ and decays as $(1+z)^{\al_i}$ for $z<z_i$ which represents a fluid with constant EoS of $(\al_i/3-1)$. For matter like behaviour we have $\al_i=3$ while for radiation like behaviour $\al_i=4$. For CC $\al_i=0$. For stiff nature with $\rho_\phi\sim a^{-6}$ we have $\al_i=6$. So, while $\al_i$ determines rate of decay of $\rho_\phi$ the other parameter $z_i$ determines the transition redshift at which the function $\rho_i(z)$ starts decaying. Now, using multiple $\rho_i(z)$ we can  reproduce different natures of $\rho_\phi$ at different epochs by choosing different $\al_i$ and $z_i$, {\it i.e.}, we can choose a function $\sum_i \rho_i(z)$. Considering this, we introduce the following parametrization of the scalar field energy density
\begin{eqnarray}
    \rho_\phi(z) &=& \sum_{i=1}^{f} \frac{\rho_{0i}}{1+\(\frac{1+z_{i}}{1+z}\)^{\al_i}}+\rho_{\rm KE} \(1+z\)^6 \, .~~~~
    \label{eq:rho_para}
\end{eqnarray}
$\rho_{\rm KE}$ sets the initial value of $\rho_\phi$. $\rho_{\rm KE}=0$ implies that the scalar field is initially frozen {\it i.e.}, its kinetic term is almost zero. If $\rho_{\rm KE}\neq 0$ then $\rho_\phi$ initially falls as $a^{-6}$ and becomes subdominant and 
because of the Hubble damping the scalar field freezes to evolve. During the frozen period $\rho_\phi$ increases and becomes comparable to the background energy density and starts evolving again. $f$ is the number of frozen periods in the scalar field dynamics which will depend on the nature of the scalar field dynamics governed by the nature of the potential, more precisely the nature and the values of the functions $\lam$ and $\Gamma$. So, $f=1$ is the first frozen period of the scalar field in the past. $\rho_{0i}$'s and $z_i$'s are constants. Even though we may need more than one $z_i$ but all the $z_i$'s, for $i>1$, can be represented in terms of $z_{i=1}$ and $\rho_{0i}$'s as $\rho_i(z_i)=\rho_{i-1}(z_i)$ using the following relation
\begin{eqnarray}
    1+z_{i>1}=\(\frac{2\rho_{0(i-1)}}{\rho_{0i}}-1\)^{-1/\al_{i-1}} (1+z_{i-1}) \, .
    \label{eq:z_val}
\end{eqnarray}
The Hubble parameter is given by the Firedmann equation
\begin{eqnarray}
    3H^2(z)\Mpl^2=\rho_\m(z)+\rho_\r(z)+\rho_\phi(z) \, ,
\end{eqnarray}
Since the total density parameter should be one, at $z=0$, we have the constraint equation
\begin{eqnarray}
    \Om_{\m0}+\Om_{\r0}+\sum_{i=1}^{f} \frac{\Om_{0i}}{1+\(1+z_{i}\)^{\al_i}}+\Om_{\rm KE} =1
\end{eqnarray}
which reduces one more free parameter. In the last equation we have considered $\Om_\phi\to\rho_\phi/\rho_{{\rm c}0}$, {\it i.e.}, $\Om_{0i}\to\rho_{0i}/\rho_{{\rm c}0}$ and $\Om_{\rm KE}\to\rho_{\rm KE}/\rho_{{\rm c}0}$. Also, $\rho_{\rm KE}$ fixes the initial energy density of the scalar field and should be very small and it is sufficient to fix the value of $\Om_{\rm KE}$ and not consider it as a model parameter as it will have almost no effect on the late time evolution of the scalar field energy density. So, the constraint equation becomes
\begin{eqnarray}
    \Om_{\m0}+\Om_{\r0}+\sum_{i=1}^{f} \frac{\Om_{0i}}{1+\(1+z_{i}\)^{\al_i}} =1
\end{eqnarray}

Now, practically, at $z=0$, the dominant contribution will come from the $i=f$ term as long as $z_{f-1}$ is not very close to 0. This gives us
\begin{eqnarray}
    \Om_{\m0}+\Om_{\r0}+\frac{\Om_{0f}}{1+\(1+z_{f}\)^{\al_f}} =1 \, .
    \label{eq:Om_Cons}
\end{eqnarray}
For $\al_f \neq 0$, Eq.~\eqref{eq:Om_Cons} gives us
\begin{eqnarray}
    1+z_{f} &=& \Om_\delta^{1/\al_f}\, , ~~~{\rm where,}   \\ 
    \Om_\delta &=& \frac{\Om_{0f}}{\Om_{\rm DE0}}-1 \, ,~~~{\rm and} \\
    \Om_{\rm DE0} &=& 1-\Om_{\m0}-\Om_{\r0} \, .
    \label{eq:z_val_rho}
\end{eqnarray}
Here, $\Om_{\rm DE0}$ represents the present value of dark energy density parameter and $\Om_\delta$ determines the value of $z_f$. When $\al_f=0$ the late time is governed by the CC and we do not need to incorporate the parameter $z_f$. So, the parameter $z_{f-1}$ can be represented in terms of the other parameters in this case $\Om_{0f}/2=\Om_{\rm DE0}$. Using the above equations we can see from Eq.~\eqref{eq:z_val} that all the $z_{0i}$'s can be represented by $\rho_{0i}$'s which reduces one more free parameter. This will be clearer in the forthcoming sections where we give explicit examples of scalar field dynamics and how the parametrization~\eqref{eq:rho_para} can reproduce the behaviour of the scalar field dynamics. So, finally, we have only $\rho_{0i}$'s as the free parameters in the parametrization~\eqref{eq:rho_para}. In the next sections, we will see that the maximum value of $i$, {\it i.e.}, $i_{\rm max}\leq 2$ is sufficient to represent the scalar field dynamics. So, we can represent scalar field dynamics with at most $4$ additional parameters, two for $\rho_{0i}$'s and two for $\al_i$'s. Eq.~\eqref{eq:Om_Cons} may not be valid when $z_{f-1}$ is very close to 0. This can happen when the kinetic energy of the scalar field still contributes significantly even at present. This scenario can particularly be observed for tracker dynamics in some potentials such as the inverse power law potential \cite{Ratra:1987rm,Steinhardt:1999nw} which does not produce a viable cosmology \cite{Steinhardt:1999nw,Hossain:2023lxs}. The inverse axionlike potential gives rise to viable cosmology along with tracker dynamics \cite{Hossain:2023lxs}. For this case the potential energy dominates from the recent past and we can use Eq.~\eqref{eq:Om_Cons} safely. We shall discuss this in details in Sec.~\ref{sec:tracker}.

The EoS of the scalar field $w_\phi$ can be parameterized from Eq.~\eqref{eq:rho_para} using the continuity equation of the scalar field
\begin{eqnarray}
    \dot{\rho}_\phi+3H\rho_\phi(1+w_\phi)=0 \, ,
    \label{eq:cont}
\end{eqnarray}
and is given by
\begin{eqnarray}
    w_\phi(z) &=& -1+\frac{w_1(z)}{3\rho_{\phi}(z)} \, . 
    \label{eq:w_para}
\end{eqnarray}
where
\begin{eqnarray}
    w_1(z) &=& -\frac{\dot\rho_\phi}{H} = \sum_{i=1}^{f}\frac{\rho_{0i}\al_i\(\frac{1+z_i}{1+z}\)^{\al_i}}{\(1+\(\frac{1+z_i}{1+z}\)^{\al_i}\)^2}+6\rho_{\rm KE} \(1+z\)^6\, ,
\end{eqnarray}

\section{Scaling-freezing dynamics}
\label{sec:scaling-freezing}

For scaling-freezing dynamics we consider the following double exponential potential \cite{Barreiro:1999zs,Jarv:2004uk}
\begin{eqnarray}
    V(\phi)=V_1\e^{-\lam_1\phi/\Mpl}+V_2\e^{-\lam_2\phi/\Mpl}\, ,
    \label{eq:potdexp}
\end{eqnarray}
for which the corresponding 
\begin{eqnarray}
    \Gamma=1-\frac{(\lam-\lam_1)(\lam-\lam_2)}{\lam^2}\, ,
    \label{eq:gam_dexp}
\end{eqnarray}
where $V_1$ and $V_2$ are constant and they set the energy scales while $\lam_1$ and $\lam_2$ are the slopes of the two exponential functions and are constants too. This potential was introduced in \cite{Barreiro:1999zs} for $V_1=V_2$. Now, considering $V_1\neq V_2$ we can relate two energy scales dynamically by a scalar field. Let's consider $V_1>V_2$. Now, to get scaling solution the slope of the exponential function associated with $V_1$ should be $>\sqrt{3}$ {\it i.e.}, $\lam_1>\sqrt{3}$ \cite{Copeland:1997et}. In fact, the fixed point associated with the scaling solution tells us that $\Om_\phi=4/\lam_1^2$ during radiation era and $\Om_\phi=3/\lam_1^2$ during matter era \cite{Copeland:1997et}. Since $\Om_\phi$ should be very small during matter era we can say, to estimate, that $\Om_\phi<0.1$ which tells us that $\lam_1>\sqrt{30}$. So, by fixing $\lam_1>\sqrt{30}$ we can get scaling solutions and maintain scalar field energy density as the small fraction in total energy density during matter era. Now, the slope $\lam$ can vary between $\lam_1$ and $\lam_2$. As long as $\lam\approx\lam_1$ we have scaling dynamics as, from Eq.~\eqref{eq:gam_dexp}, we see $\Gamma\approx 1$ for this case and we have considered large values of $\lam_1$. Once $\lam$ starts moving away from $\lam_1$ the function $\Gamma$ also starts moving away from $1$ which results to the exit from the scaling behaviour in the scalar field dynamics. Eventually $\lam$ becomes same as $\lam_2$ and $\Gamma$ again becomes 1 but if we consider $\lam_2<\sqrt{3}$ then we don't have scaling solution rather we can have late time acceleration. So, to get late time acceleration we can fix $V_2$ at the dark energy scale with small value of $\lam_2<\sqrt{3}$. One should also note that, if we consider negative values of $\lam_2$ then the potential~\eqref{eq:potdexp} becomes an oscillatory potential which can also give rise to late time acceleration \cite{Turner:1983he}. 

\begin{figure}[t]
\centering
\includegraphics[scale=.93]{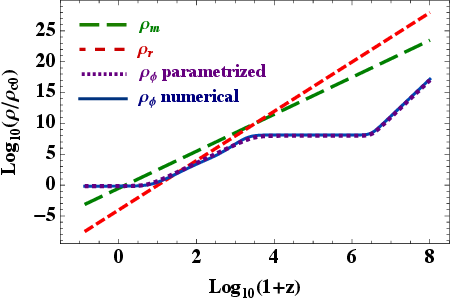} ~~~
\includegraphics[scale=.94]{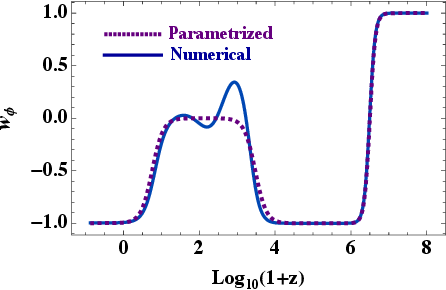} \vskip10pt
\includegraphics[scale=.93]{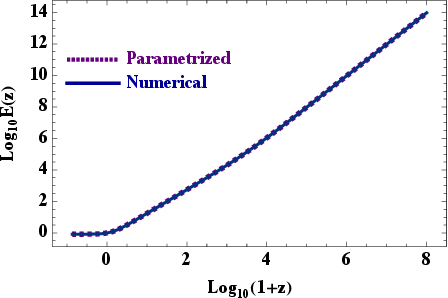} ~~~
\includegraphics[scale=.94]{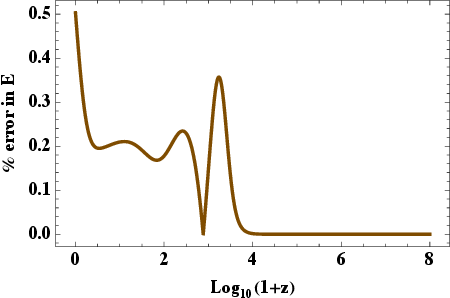} \vskip10pt
\includegraphics[scale=.8]{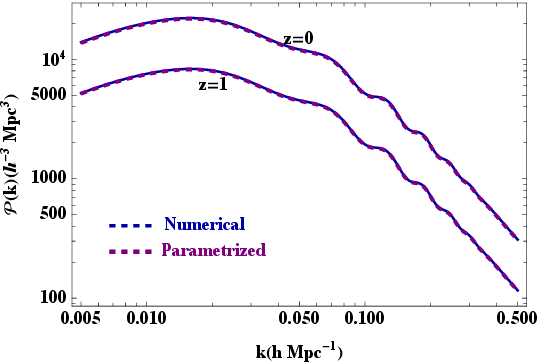} ~~~
\includegraphics[scale=.8]{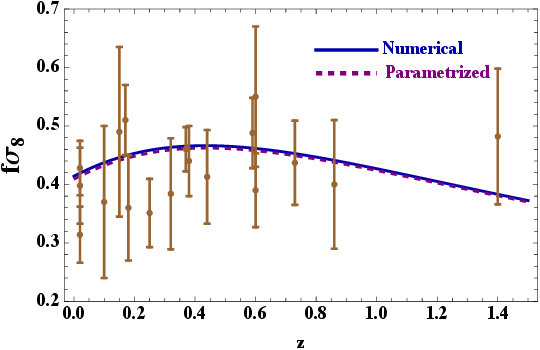} 
\caption{(Top left) Evolution of energy densities of matter (long dashed green), radiation (short dashed red) and scalar field (solid blue line is numerically evolved and dotted purple line is parametrized) normalised with the present value of critical density $\rho_{\rm c0}$ along with the numerically evolved and parametrized EoS of the scalar field (top right), normalised Hubble parameter $E(z)$ (middle left), the percentage error in $E(z)$ (middle right), (bottom left) matter power spectrum for $z=0$ (upper) and $z=1$ (lower) and (bottom right) evolution of $f\sig_8$ along with the observational data and their error bars have been shown. in the bottom right figure the brown dots are the observational data of $f\sig_8(z)$ along with their $1\sig$ error bars \cite{Nesseris:2017vor}. For the numerical curves the initial conditions and parameter values are $V_1=10^9\rho_{\rm c0}$, $V_2=0.79\rho_{\rm c 0}$, $\lam_1=20$, $\lam_2=0.1$, initial field value $\phi_i=0.1 \Mpl$ and $\phi_i'=\d\phi_i/\d \ln(1+z)=10^{-5}\Mpl$. For the parametrized curves we have taken $\Om_{\rm KE}=10^{-31}$, $\Om_{01}=10^8$, $\Om_\delta=1.018$ ($\Om_{02}=1.4123,\; z_2=4.858$), $\al_1=3$  and $\al_2=0.01$.} 
\label{fig:rho_para_dexp}
\end{figure}

Considering that the scaling behaviour exists during the matter era then for scaling-freezing dynamics we can consider $f=2$ and the parametrization~\eqref{eq:rho_para} becomes
\begin{eqnarray}
    \rho_\phi(z) = \frac{\rho_{02}}{1+\(\frac{1+z_{2}}{1+z}\)^{\al_2}}+\frac{\rho_{01}}{1+\(\frac{1+z_1}{1+z}\)^{\al_1}}+\rho_{\rm KE} \(1+z\)^6 \, .
    \label{eq:rho_para_scaling}
\end{eqnarray}
For this case Eq.~\eqref{eq:z_val} reduces to
\begin{eqnarray}
    1+z_2=\(\frac{2\rho_{01}}{\rho_{02}}-1\)^{-1/\al_1} (1+z_1)
\end{eqnarray}
and Eq.~\eqref{eq:z_val_rho} reduces to (using $\Om_{0i}=\rho_{0i}/\rho_{\rm c0}$)
\begin{eqnarray}
    1+z_2 = \(\frac{\Om_{02}}{\Om_{\rm DE0}}-1\)^{1/\al_2} = \Om_\delta^{1/\al_2} \, .
    \label{eq:z2_sc}
\end{eqnarray}
From the last two equations we have 
\begin{eqnarray}
    1+z_1 = \(\frac{\Om_{02}}{\Om_{\rm DE0}}-1\)^{1/\al_2} \(\frac{2\Om_{01}}{\Om_{02}}-1\)^{1/\al_1} = \Om_\delta^{1/\al_2} \(\frac{2\Om_{01}}{(1+\Om_\delta)\Om_{\rm DE0}}-1\)^{1/\al_1}
    \label{eq:z1_sc}
\end{eqnarray}
So, the extra parameters are $\Om_\delta$, $\Om_{01}$, $\al_1$ and $\al_2$. $\rho_{\rm KE}$ sets the initial value of $\rho_\phi$. $\rho_{\rm KE}=0$ implies that the scalar field is initially frozen {\it i.e.}, its kinetic term is almost zero. If $\rho_{\rm KE}\neq 0$ then $\rho_\phi$ initially falls as $a^{-6}$ and becomes subdominant which causes huge Hubble damping. 
Because of this Hubble damping the scalar field freezes to evolve. During the frozen period $\rho_\phi$ increases and becomes comparable to the background energy density and starts evolving again. This redhsift from which the scalar field starts evolving again is denoted by $z_1$ and scalar field energy density at $z_1$ is associated with $\rho_{01}$. If the potential is very steep then $\rho_\phi$ may again fall as $a^{-6}$ and repeat the previous dynamics. Since we have considered that the scalar field scales the matter we can safely choose $\al_1=3$. If the scaling behaviour starts from radiation era we have to consider $f=3$. For $z_2\leq z \leq z_1$ the scaling regime persists and for $z<z_2$ the scalar field starts deviating from the scaling regime. If we choose $\al_2<1$ the scalar field EoS will be closer to $-1$ at $z=0$. The EoS of the scalar field follows from Eq.~\eqref{eq:w_para}
\begin{eqnarray}
    w_\phi(z) &=& -1+\frac{1}{3\rho_\phi} \(\frac{\rho_{02}\al_2\(\frac{1+z_2}{1+z}\)^{\al_2}}{\(1+\(\frac{1+z_2}{1+z}\)^{\al_2}\)^2}+\frac{\rho_{01}\al_1\(\frac{1+z_1}{1+z}\)^{\al_1}}{\(1+\(\frac{1+z_1}{1+z}\)^{\al_1}\)^2}+6\rho_{\rm KE} \(1+z\)^6\) \, ,
    \label{eq:eos_f2}
\end{eqnarray} 
At $z=0$, ignoring the last term in Eqs.~\eqref{eq:rho_para_scaling} and \eqref{eq:eos_f2} we have 
\begin{eqnarray}
    w_\phi(0) = w_0 = -1+\frac{1}{3\(\Om_{\rm DE0}+\frac{\Om_{01}}{1+(1+z_1)^{\al_1}}\)}\(\frac{\al_2\Om_{\rm DE0}\Om_\delta}{1+\Om_\delta}+\frac{\al_1\Om_{01}(1+z_1)^{\al_1}}{(1+(1+z_1)^{\al_1})^2}\) \, .
    \label{eq:w0_sc}
    \end{eqnarray}

Fig.~\ref{fig:rho_para_dexp} compares the numerical results with the parametrization~\eqref{eq:rho_para_scaling} for the double exponential potential~\eqref{eq:potdexp}. We can see that the parametrization, represented by the dotted purple lines, mimics the numerical results, represented by solid blue lines. Around the redshift range $100$ to $1000$ the parametrized EoS, given in Eq.~\eqref{eq:eos_f2} gives the average value of the numerically evolved EoS which has a oscillatory behaviour during that period (top right figure of Fig.~\ref{fig:rho_para_dexp}). We can match this oscillatory behaviour by increasing one more value of $f$ which we will show shortly. The evolution of the normalized Hubble parameter $E(z)=H(z)/H_0$ is same for both parametrized and numerically evolved cases (middle left figure of Fig.~\ref{fig:rho_para_dexp}). The similarity in the evolution in $E(z)$ has been quantified in the middle right figure of Fig.~\ref{fig:rho_para_dexp} that shows the {\it percentage error in} $E(z)=\frac{|E_{\rm parametrized}(z)-E_{\rm numerical}(z)|}{E_{\rm numerical}(z)}\times 100 \%$ and the maximum error that we get is around $0.4\%$. This shows that the amount of mismatch we have in the evolution of $w_\phi$ does not have any effect on the evolution of $E(z)$ and therefore the parametrization should lead to observational predictions similar to the numerical results. The top and middle figures of Fig.~\ref{fig:rho_para_dexp} show the validity of the parametrization~\eqref{eq:rho_para_scaling} at the background level. To check the consistency of the parametrization with the numerical results at the perturbation level we have shown the matter power spectrum and the evolution of $f\sig_8(z)$ in the bottom left and bottom right figures of Fig.~\ref{fig:rho_para_dexp} respectively. We see that the parametrization~\eqref{eq:rho_para_scaling} predicts the power spectrum and the evolution of $f\sig_8(z)$ similar to the numerical results. So, Fig.~\ref{fig:rho_para_dexp} shown the validity of the parametrization~\eqref{eq:rho_para_scaling} not only at the background cosmology level but also at the perturbation level.

\begin{figure}[t]
\centering
\includegraphics[scale=.94]{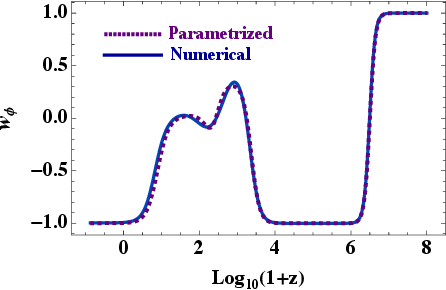} ~~~
\includegraphics[scale=.93]{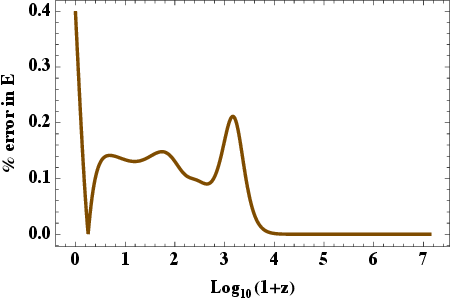} 
\caption{Evolution of scalar field EoS (left) and the percentage error in $E(z)$ has been shown. For the numerical curves the initial conditions and parameter values are $V_1=10^9\rho_{\rm c0}$, $V_2=0.79\rho_{\rm c 0}$, $\lam_1=20$, $\lam_2=0.1$, initial field value $\phi_i=0.1 \Mpl$ and $\phi_i'=\d\phi_i/\d \ln(1+z)=10^{-5}\Mpl$. For the parametrized curves we have taken $\Om_{\rm KE}=10^{-31}$, $\Om_{01}=10^8$, $\Om_{02}=1.3\times 10^4$, $\Om_\delta=1.021$ ($\Om_{03}=1.4145,\; z_3=6.973$), $\al_1=4$, $\al_2=3$ and $\al_3=0.01$.}. 
\label{fig:rho_para_dexp_m3}
\end{figure}

As we already have mentioned that the mismatch in the evolution of $w_\phi$ in the numerical and parametrized results, during the redshift range $100$ to $1000$, can be reduced by considering $f=3$ instead of considering $f=2$. For $f=3$ the parametrization becomes
\begin{eqnarray}
    \rho_\phi(z) = \frac{\rho_{03}}{1+\(\frac{1+z_{3}}{1+z}\)^{\al_3}}+\frac{\rho_{02}}{1+\(\frac{1+z_{2}}{1+z}\)^{\al_2}}+\frac{\rho_{01}}{1+\(\frac{1+z_1}{1+z}\)^{\al_1}}+\rho_{\rm KE} \(1+z\)^6 \, ,
    \label{eq:rho_para_scaling_m3}
\end{eqnarray}
and the relations between the parameters will be
\begin{eqnarray}
    1+z_3 &=& \(\frac{\Om_{03}}{\Om_{\rm DE0}}-1\)^{1/\al_3} = \Om_\delta^{1/\al_3} \, . \\
    1+z_2 &=&  \Om_\delta^{1/\al_3} \(\frac{2\Om_{02}}{(1+\Om_\delta)\Om_{\rm DE0}}-1\)^{1/\al_2} \, , \\
    1+z_1 &=&   \Om_\delta^{1/\al_3} \(\frac{2\Om_{02}}{(1+\Om_\delta)\Om_{\rm DE0}}-1\)^{1/\al_2} \(\frac{2\Om_{01}}{\Om_{02}}-1\)^{1/\al_1}\, .
\end{eqnarray}
As we choose $f=3$ the parametrization almost mimics the evolution of $w_\phi$ which is shown in the left figure of Fig.~\ref{fig:rho_para_dexp_m3}. Although we have this improvement in the evolution of $w_\phi$, it has almost no effect on the evolution of $E(z)$ which has been shown in the right figure of Fig.~\ref{fig:rho_para_dexp_m3}. So, we can say that the parametrization~\eqref{eq:rho_para_scaling} is sufficient to mimic any scaling-freezing dynamics.

\section{Tracker dynamics}
\label{sec:tracker}

\begin{figure}[t]
\centering
\includegraphics[scale=.93]{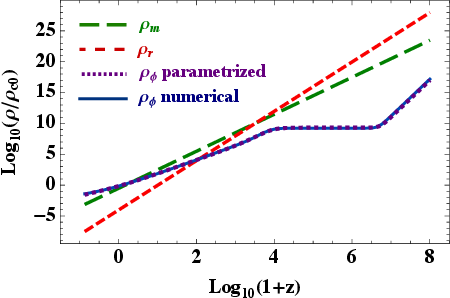} ~~~
\includegraphics[scale=.94]{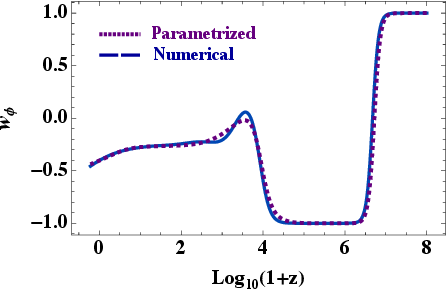} \vskip10pt
\includegraphics[scale=.93]{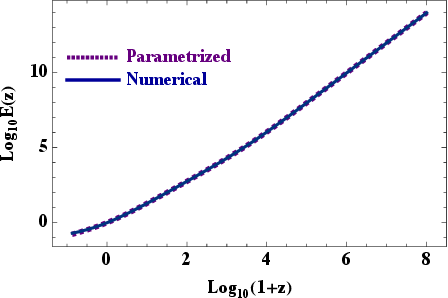} ~~~
\includegraphics[scale=.94]{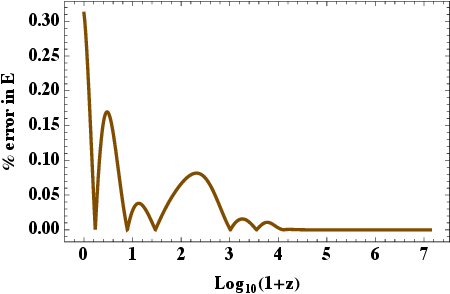} 
\caption{Figures represent the same cosmological parameters as the upper and middle figures of Fig.~\ref{fig:rho_para_dexp} for the potential~\eqref{eq:pot_pl} and parametrization~\eqref{eq:rho_para_scaling_m3}. For the numerical curves the initial conditions and parameter values are $V_0=1700\rho_{\rm c0}$,  $n=6$, initial field value $\phi_i=0.1 \Mpl$ and $\phi_i'=\d\phi_i/\d \ln(1+z)=10^{-5}\Mpl$. For the parametrized curves we have taken $\Om_{\rm KE}=10^{-31}$, $\Om_{01}=1.5\times 10^9$, $\Om_{02}=5.7\times 10^8$, $\Om_{03}=1.1$, $z_1=9000$, $\al_1=3.5$, $\al_2=2.2$ and $\al_3=1.3$.}. 
\label{fig:rho_para_track_pl}
\end{figure}

\begin{figure}[t]
\centering
\includegraphics[scale=.93]{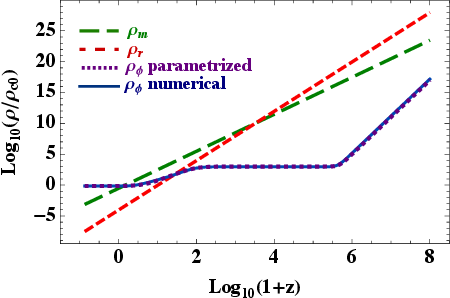} ~~~
\includegraphics[scale=.94]{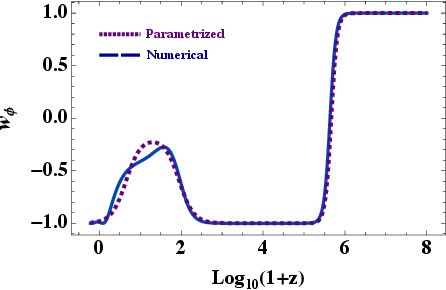} \vskip10pt
\includegraphics[scale=.93]{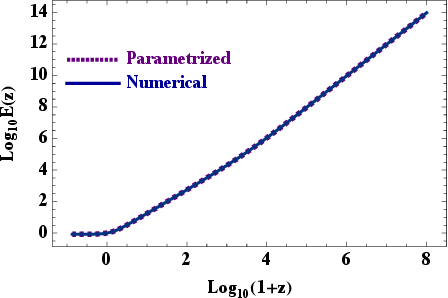} ~~~
\includegraphics[scale=.94]{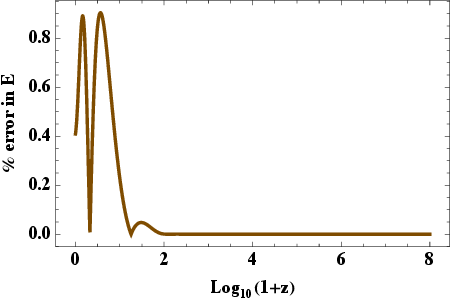} \vskip10pt
\includegraphics[scale=.8]{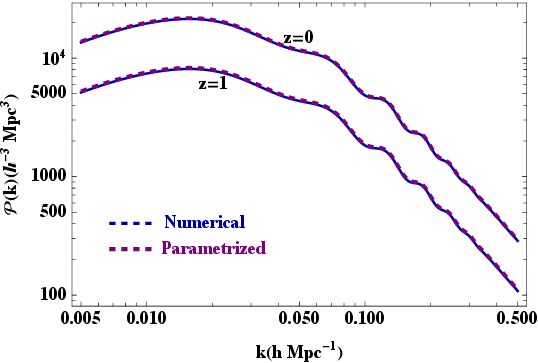} ~~~
\includegraphics[scale=.8]{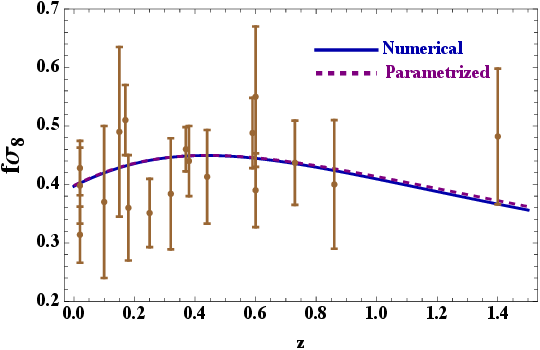} 
\caption{Curves represent the same cosmological parameters as Fig.~\ref{fig:rho_para_dexp} for the potential~\eqref{eq:pot_iax} and parametrization~\eqref{eq:rho_para_scaling}. For the numerical evolution we have considered $n=2$ and the initial conditions are $\phi_i=0.1 \Mpl$ and $\phi_i'=\d\phi_i/\d \ln(1+z)=10^{-5}\Mpl$. For the parametrized curves we have taken $\Om_{\rm KE}=10^{-31}$, $\Om_{01}=10^3$, $z_1=88$ and $\al_1=2.44$.}. 
\label{fig:rho_para_track_iax}
\end{figure}

For tracker dynamics we consider the inverse power law potential \cite{Ratra:1987rm,Steinhardt:1999nw}
\begin{eqnarray}
    V(\phi)=V_0\(\frac{\Mpl}{\phi}\)^{n} \, 
    \label{eq:pot_pl}
\end{eqnarray}
and the inverse axionlike potential \cite{Hossain:2023lxs}
\begin{eqnarray}
    V(\phi)=V_0\(1-\cos\(\frac{\phi}{f_{\rm pl}}\)\)^{-n} \, ,
    \label{eq:pot_iax}
\end{eqnarray}
where, $n$, $f_{\rm pl}$, $V_0$ are constants and $n>0$. For $n<0$ the potential~\eqref{eq:pot_iax} becomes the axionlike potential which is a well studied potential in cosmology \cite{Marsh:2015xka,Poulin:2018cxd,Poulin:2018dzj,Poulin:2023lkg,Smith:2019ihp,Hlozek:2014lca}. In tracker dynamics, the scalar field energy density $\rho_\phi$ has a frozen period during early times. After the frozen period $\rho_\phi$ does not exactly follow the background and instead decays, at least during late times, a little slower than the background. Because of this nature the scalar field energy density takes over the matter but if the potential does not have necessary shallow region the kinetic energy of the scalar field may contribute significantly. This results in a much larger EoS than $-1$. This happens for the inverse power potential~\eqref{eq:pot_pl} and we can't get viable cosmology in this case if we have tracker dynamics \cite{Steinhardt:1999nw,Hossain:2023lxs}. This problem does not exist in the inverse axionlike potential~\eqref{eq:pot_iax} as the potential can generate a CC like term during the late time \cite{Hossain:2023lxs} which leads to a viable cosmology. This special feature of this potential also relates the dark energy scale with any higher energy scale. In other words, we can generate CC like term from any higher energy scale by tuning the parameters of the potential \cite{Hossain:2023lxs}.

Tracker dynamics can be parametrized with the parametrization~\eqref{eq:rho_para_scaling_m3}. For the inverse power law potential~\eqref{eq:pot_pl}, as we have significant contribution from kinetic energy even during the late times, the constraint equation~\eqref{eq:Om_Cons} changes to
\begin{eqnarray}
    \Om_{\m0}+\Om_{\r0}+\frac{\Om_{0f}}{1+\(1+z_{f}\)^{\al_f}} +\frac{\Om_{0(f-1)}}{1+\(1+z_{f-1}\)^{\al_{f-1}}} =1 \, ,
    \label{eq:Om_Cons_track}
\end{eqnarray}
which, for the parametrization~\eqref{eq:rho_para_scaling_m3}, becomes
\begin{eqnarray}
    \Om_{\m0}+\Om_{\r0}+\frac{\Om_{03}}{1+\(1+z_{3}\)^{\al_3}} +\frac{\Om_{02}}{1+\(1+z_2\)^{\al_2}} =1 \, ,
    \label{eq:Om_Cons_track1}
\end{eqnarray}
In the above equation we have neglected the term associated with $\Om_{01}$ as this term is significant only during the high redshift and becomes insignificant around $z=0$. From above equation we can represent $z_3$ in terms of the other parameters. Now, $z_2$ can be represented, following Eq.~\eqref{eq:z_val}, as
\begin{eqnarray}
    1+z_2&=&\(\frac{2\Om_{01}}{\Om_{02}}-1\)^{-1/\al_1}(1+z_1)\, . 
\end{eqnarray}
So, in this case we have to consider $z_1$ as a free parameter. Figs.~\ref{fig:rho_para_track_pl} shows the comparison among the numerical results and parametrized results using the parametrization~\eqref{eq:rho_para_scaling_m3} for the inverse power law potential~\eqref{eq:pot_pl}. We can see that the parametrization~\eqref{eq:rho_para_scaling_m3} mimics the tracker dynamics achieved numerically. The maximum difference in the evolution of the Hubble parameter is about 0.3$\%$.

For the inverse axionlike potential~\eqref{eq:pot_iax} we can relate $V_0$ with the dark energy density as \cite{Hossain:2023lxs}
\begin{eqnarray}
    V_0/\rho_{\rm c 0}=2^n \Om_{\rm DE0}\, ,
\end{eqnarray}
{\it i.e.}, we get a CC like term automatically in this potential during the late time. So, in this potential the dynamics around $z=0$ is very similar to the standard $\Lambda$CDM model. Considering this we can choose $f=2$ with $\al_2=0$. So, the parametrization can be
\begin{eqnarray}
    \rho_\phi(z) = \frac{\rho_{02}}{2}+\frac{\rho_{01}}{1+\(\frac{1+z_1}{1+z}\)^{\al_1}}+\rho_{\rm KE} \(1+z\)^6 \, ,
    \label{eq:rho_para_track_iax}
\end{eqnarray}
where $\rho_{02}=2\Om_{\rm DE0}$. If we fix $\rho_{\rm KE}$ then the free parameters are $\rho_{01}$, $z_1$ and $\al_1$. In Fig.~\ref{fig:rho_para_track_iax} the comparison between the numerical and parametrized results has been shown in terms of different cosmological parameters. We can see that the parametrization~\eqref{eq:rho_para_track_iax} fits quiet well with the numerical results for both at the background and perturbation level.
\section{Thawing dynamics}
\label{sec:thawing}

\begin{figure}[t]
\centering
\includegraphics[scale=.93]{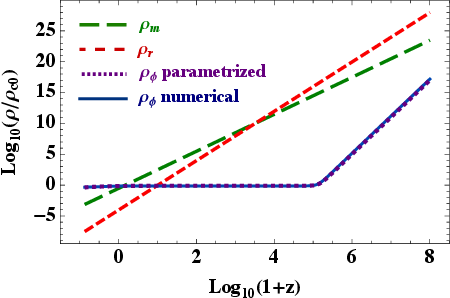} ~~~
\includegraphics[scale=.94]{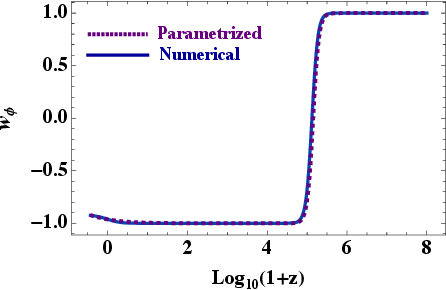} \vskip10pt
\includegraphics[scale=.93]{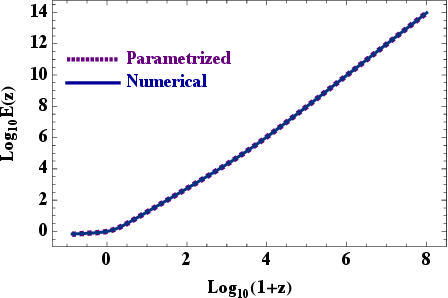} ~~~
\includegraphics[scale=.94]{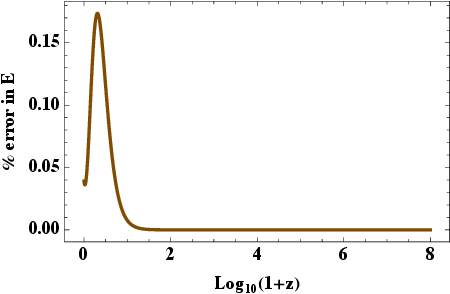} \vskip10pt
\includegraphics[scale=.8]{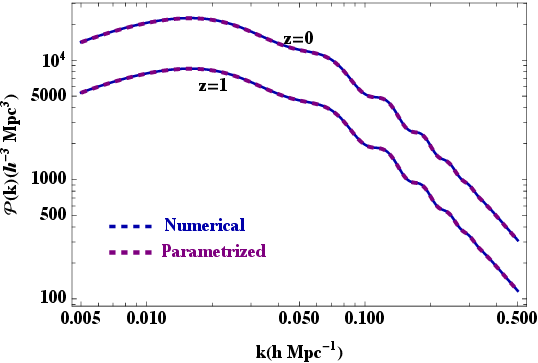} ~~~
\includegraphics[scale=.8]{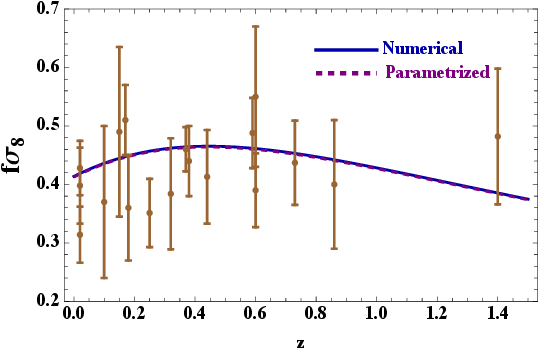} 
\caption{Similar figure as Fig.~\ref{fig:rho_para_dexp} have been shown for the potential~\eqref{eq:potexp}. For the numerical curves the initial conditions and parameter values are $V_0=0.95\rho_{\rm c0}$,  $\beta=0.5$, initial field value $\phi_i=0.5 \Mpl$ and $\phi_i'=\d\phi_i/\d \ln(1+z)=10^{-5}\Mpl$. For the parametrized curves we have taken $\Om_{\rm KE}=10^{-31}$, $\Om_\delta=0.143$ ($\Om_{01}=0.8,\; z_1=-0.8847$), $\al_1=0.9$.}
\label{fig:rho_para_thaw}
\end{figure}

In thawing dynamics the scalar field remains frozen for most of the time and starts rolling down the potential from the recent past. So the dynamics is very similar to the standard $\Lambda$CDM except at the present time when the EoS can deviate from $-1$. So, for thawing dynamics we can choose $f=1$ and the parametrization of the scalar field energy density becomes
\begin{eqnarray}
    \rho_\phi(z) = \frac{\rho_{01}}{1+\(\frac{1+z_1}{1+z}\)^{\al_1}}+\rho_{\rm KE} \(1+z\)^6 \, ,
    \label{eq:rho_para_thaw}
\end{eqnarray}
and using Eq.~\eqref{eq:z_val_rho} we have
\begin{eqnarray}
     1+z_1 &=& \(\frac{\Om_{01}}{\Om_{\rm DE0}}-1\)^{1/\al_1} = \Om_\delta^{1/\al_1}\, .
     \label{eq:z1_thaw}
\end{eqnarray}
The EoS corresponding to the parametrization~\eqref{eq:rho_para_thaw} is
\begin{eqnarray}
    w_\phi(z) &=& -1+\frac{1}{3\rho_\phi} \(\frac{\rho_{01}\al_1\(\frac{1+z_1}{1+z}\)^{\al_1}}{\(1+\(\frac{1+z_1}{1+z}\)^{\al_1}\)^2}+6\rho_{\rm KE} \(1+z\)^6\) \, ,
    \label{eq:eos_thaw}
\end{eqnarray} 
At $z=0$, by ignoring the last term in Eqs.~\eqref{eq:rho_para_thaw} and \eqref{eq:eos_thaw}, we get
\begin{eqnarray}
    w_\phi(0) = w_0 = -1+\frac{\al_1}{3}\frac{\Om_\delta}{1+\Om_\delta}
    \label{eq:w0}
\end{eqnarray}
Using the above equation we can use $w_0$ as a model parameter instead of $\al_1$. It should also be noted that the parametrization~\eqref{eq:rho_para_thaw} can accommodate phantom models for either $\al_1<0$ or $-1<\Om_\delta<0$. 

For the numerical purpose we consider an exponential potential of the following form
\begin{eqnarray}
    V(\phi)=V_0\e^{-\bet \phi/\Mpl}\, .
    \label{eq:potexp}
\end{eqnarray}

Fig.~\ref{fig:rho_para_thaw} compares the numerical results with the parametrization~\eqref{eq:rho_para_thaw} for the exponential potential~\eqref{eq:potexp} for thawing dynamics. We can see that the parametrization mimics the numerical results. The similarity in the evolution has been quantified in the lower right figure of Fig.~\ref{fig:rho_para_thaw} which shows the {\it percentage error in} $E(z)$ and the maximum error that we are getting is around $0.15\%$. The lower figures of Fig.~\ref{fig:rho_para_thaw} shows the power spectrum (left) and the evolution of $f\sig_8(z)$ (right) for both numerical and parametrized cases and we can see that the parametrized results mimic the numerical ones.

\section{Observational Constraints}
\label{sec:obs}
In this section we study the observational constraints on the standard $\Lambda$CDM model, $w$CDM, CPL parametrization and models with the parametrization~\eqref{eq:rho_para} with $f=1$ and $f=2$. We also compare these models by calculating the Akaike information criterion (AIC) and Bayesian Information Criterion (BIC) \cite{Akaike74,Liddle:2006tc,Trotta:2017wnx,Shi:2012ma} along with the minimum chi-squared ($\chi_{\rm min}^2$) and reduced chi-squared ($\chi_{\rm red}^2=\chi_{\rm min}^2/\nu$), where $\nu=k-N$ is the degree of freedom while $k$ and $N$ are the total number of data points and the total number of model parameters respectively. AIC and BIC are defined as 
\begin{eqnarray}
   \rm AIC &=& 2N - 2\ln\mathcal{L}_{\rm max} 
 = 2N + \chi^2_{\rm min} \, , \\
   \rm BIC &=& N \ln K - 2\ln\mathcal{L}_{\rm max} = N \ln K + \chi_{\rm min}^2 \, ,
\end{eqnarray}
where, $\mathcal{L}_{\rm max}$ is the maximum likelihood. We also compute the difference in AIC ($\Delta$AIC) and BIC ($\Delta$BIC) between the parametrized models with $f=1$ and $f=2$ and $\Lambda$CDM model such that
\begin{eqnarray}
    \Delta \rm AIC &=& \rm AIC_{\rm para} - AIC_{\rm \Lambda CDM}  \\ 
    \Delta \rm BIC &=& \rm BIC_{\rm para} - BIC_{\rm \Lambda CDM} \, ,
\end{eqnarray}
where, the subscript $para$ stands for $parametrized$. $\Delta$AIC and $\Delta$BIC tell us about the preference of the model by the observational data in comparison to a reference model which, in this case, we have considered $\Lambda$CDM model which also has less number of free parameters than the parametrized models.

For $\Lambda$CDM model  we have six parameters $\{\Om_{\m0}, h, \omega_{\rm b}, r_{\rm d}h, \sig_8, M\}$, $\omega_{\rm b}=\Om_{\m0}h^2$, $r_\d$ is the sound horizon at the decoupling and $M$ is the absolute magnitude. We consider uniform priors of the parameters as
\begin{eqnarray}
    \{\Om_{\m0},\; h,\; \omega_{\rm b},\; r_{\rm d}h,\; \sig_8,\; M\} \equiv&& \{[0.2,0.5],\; [0.5,0.8],\; [0.005,0.05],\; [60, 140],\; [0.5,1],  [-22, -15]\} \nn  \; .
\end{eqnarray}

For $w$CDM we have only one extra parameter $w_0$, present EoS of dark energy. In $w$CDM parametrization dark energy EoS $w$ remains same as $w_0$ at any redhsift. We consider the uniform prior $w_0\equiv \{-2,1\}$. For CPL parametrization~\eqref{eq:cpl_wphi} we have two more extra parameters, $w_0$ and $w_{\rm a}$. We consider the following uniform priors for the parameters: $\{w_0,w_{\rm a}\}\equiv \{[-2,0],[-2,1]\}$.

From Eq.~\eqref{eq:rho_para_thaw} we can see that for the parametrized model with $f=1$ we have two extra parameters, $\Om_\delta$ and $\al_1$, along with other six parameters mentioned above for the $\Lambda$CDM model. We call this model as $Pf1$. Keeping the same prior for the common six parameters we choose the priors of the two extra parameter as $\{\Om_\delta,\; \al_1\} \equiv \{[0,1000],\; [0,2]\}$ to consider only the non-phantom region ($w_\phi\geq -1$). This can be clear from Eq.~\eqref{eq:w0} which tells us that $w_0\geq -1$ for the chosen priors. $\al_1\equiv\{0,2\}$ correspond to the equation of state $w=\al_1/3-1\equiv\{-1,-1/3\}$. $\Om_\delta$ along with $\al_1$ determines the value of $z_1$ as $(1+z_1)^{\al_1}=\Om_\delta$ (Eq.~\eqref{eq:z1_thaw}). $z_1=-1$ for $\Om_\delta=0$ and $\al_1\neq 0$ and becomes very large for $\Om_\delta>1$ and $\al_1\to 0$. So, our choice of priors include almost all possible values of $z_1$. In fact, since all values of $\Om_\delta>1$ can give large values of $z_1$ for $\al_1\to 0$ the constraint on $\Om_\delta>1$ can be very weak. To incorporate the phantom region we consider the prior $\{\Om_\delta,\; \al_1\} \equiv \{[-0.9,1000],\; [-2,2]\}$ and we call this model as $Pf1+Phantom$. $\Om_\delta>-1$ is chosen as from Eq.~\eqref{eq:z1_thaw} we can see that for $\Om_\delta=-1$ the parameter $\Om_{01}=0$ which makes the parametrization~\eqref{eq:rho_para_thaw} unsuitable to represent late time universe. Even though $w_0 \approx -1 + \al_1/3$ also holds for $\Om_\delta < -1$, similar to the case of $\Om_\delta > 1$, it results in a negative $\Om_{01}$. Therefore, we restrict our parameter space to $\Om_\delta > -1$.

$f=2$ corresponds to the parametrization~\eqref{eq:rho_para_scaling}. We call this model as $Pf2$. This parametrization can represent both scaling-freezing and tracker dynamics as depicted above. For this parametrization we have four extra parameter, $\Om_\delta$, $\Om_{01}$, $\al_2$ and $\al_1$. We consider the following prior for these parameters, 
\begin{eqnarray}
    \{\Om_\delta,\; \Om_{01},\; \al_2,\; \al_1\} \equiv \{[0,100],\; [0,14],\; [-0.2,2],\; [0.01,6]\} \; . \nn
\end{eqnarray}
For $Pf2$ $\Omega_\delta$ and $\al_2$ determine $z_2$ while all the four parameters determine $z_1$. Now, $z_1>z_2$. Also, for viable cosmology we have to choose the values of the parameters such a way so that $\rho_\phi$ remains subdominant until the recent past. We have numerically checked the evolution of the cosmological parameters top choose the viable priors for the parameters. $\Om_{01}$ sets the scale of the frozen period in the evolution of $\rho_\phi$ as depicted in the Fig.~\ref{fig:rho_para_dexp}. $\Om_{01}$ needs to have a non-zero value so that we get scaling-freezing or tracker dynamics. If $\Om_{01}=0$ then we can get either a CC like or thawing dynamics during late time.

We perform Markov Chain Monte Carlo (MCMC) analysis to constrain the model
parameters. We use
the publicly available code {\tt EMCEE} \cite{Foreman-Mackey:2012any} for the purpose of MCMC simulation. For analysing the results and plotting the contours of the model parameters we use another publicly available python package {\tt GetDist} \cite{Lewis:2019xzd}. For assessing chain convergence, we consider the Gelman-Rubin statistic \cite{Gelman:1992zz} according to which the chains are converged when $|R-1|\lesssim 0.01$.

\subsection{Observational data}
We consider the data from the observations of cosmic microwave background (CMB) radiation, baryon acoustic oscillation (BAO), type-Ia  supernovae (SNeIa) and redshift space distortion (RSD).
\subsubsection{CMB}
The CMB distance prior uses the positions of the acoustic peak to determine the cosmological distance at the fundamental level. This prior is commonly incorporated using the following key parameters: shift parameter(R), acoustic scale ($l_{\rm A}$). We use these distance priors reconstructed from the Planck TT,TE,EE$+$lowE data of 2018 \cite{Planck:2018vyg} and given in \cite{2019JCAP}. Along with these parameters we also consider the observational bound on the baryon energy density ($\omega_{\rm b}$).

\subsubsection{BAO data from DESI DR1} 
In DESI DR1 BAO measurements \cite{DESI:2024mwx,DESI:2024uvr,DESI:2024lzq} we have the measurements from
the galaxy, quasar and Lyman-$\al$ forest tracers within the redshift range
$0.1< z< 4.2$. These include the bright galaxy sample (BGS) within the redshift range $0.1<z<0.4$, luminous red galaxy sample
(LRG) in the redshift range $0.4<z<0.6$ and $0.6<z<0.8$, emission line galaxy sample (ELG) in $1.1<z<1.6$, combined LRG and ELG sample in  $0.8<z<1.1$, the quasar
sample (QSO) in $0.8<z<2.1$ \cite{DESI:2024uvr} and the Lyman-$\al$ Forest
Sample (Ly-$\al$) in $1.77<z<4.16$ \cite{DESI:2024lzq}.

\subsubsection{Type-Ia Supernova}
We consider the distance moduli measurements from the PantheonPlus (PP) sample of Type-Ia supernovae (SNeIa), which consists of 1550 SneIa luminosity distance measurements within  the redshift range $0.001 < z < 2.26$ \cite{Scolnic:2021amr,Brout:2022vxf}.

\subsubsection{Observational Hubble Data}
We analyse observational data for the Hubble parameter measured at various redshifts
within the redshift range of 0.07 to 1.965. We focus on a collection of 31 $H(z)$ measurements derived using the cosmic chronometric method \cite{2018JCAP...04..051G}.

\subsubsection{Redshift Space Distortion}
We consider the redshift space distortion (RSD) measurements of the cosmological growth rate, $f\sig_8(z)$, from different surveys compiled in \cite{Nesseris:2017vor}. $f(a)$ is the growth factor and defined as
\begin{eqnarray}
    f(a) = \frac{\d \ln \delta (a)}{\d \ln a} \, ,
\end{eqnarray}
where, $\delta (a)$ is the matter density contrast, $\delta \rho_\m/\rho_\m$, $\delta \rho_\m$ being the matter density fluctuation of the background matter density $\rho_\m$ and $\sig_8(z)$ is the root mean square amplitude of mass fluctuations within spheres of radius $8h^{-1}$Mpc and given by
\begin{eqnarray}
    \sig_8(z) = \sig_8 \frac{\delta (z)}{\delta (0)} \, ,
\end{eqnarray}
where, $\sig_8$ is the present value of $\sig_8(z)$, {\it i.e.}, $\sig_8(0)$. We follow \cite{Nesseris:2017vor} to construct the covariance matrix and define the $\chi^2$ for the RSD data. In this regard, it should be noted that the RSD data, from different measurements, have  a dependence on the fiducial model used by the collaborations to convert redshifts to distances. To correct this we have to define a ratio
\begin{eqnarray}
    {\rm ratio}(z) = \frac{H(z)D_A(z)}{H_{\rm fid}(z)D_{A,\rm fid}(z)} \,
    \label{eq:ratio}
\end{eqnarray}
where subscript $fid$ stands for $fiducial$ and the angular diameter distance $D_{A}(z)=(c/H_0)d_{A}(z)$, where $d_{A}(z)=(1/(1+z))\int \d z/E(z)$ with $E(z)$ being the dimensionless Hubble parameter, $H(z)/H_0$. The product $H(z)D_{A}(z)$ can be written as $E(z)d_{A}(z)$. Once we define the ratio~\eqref{eq:ratio} we can define the $\chi^2$ as
\begin{eqnarray}
    \chi_{\rm RSD}^2 = V^i C_{ij}^{-1} V^j \, ,
\end{eqnarray}
where, the vector $V^i(z_i,\theta)$ is given by
\begin{eqnarray}
    V^i(z_i,\theta) = f\sig_{8,i} - {\rm ratio}(z_i) f\sig_8(z_i,\theta) \, ,
\end{eqnarray}
where, $f\sig_{8,i}$ is the value of the $i$th data point at the redshift $z_i$. $\theta$ represents the model parameters. $C_{ij}$ is the covariance matrix which is an $N\times N$ diagonal matrix except at the positoons of WiggleZ data as except the data from WiggleZ the other data are not correlated. So the covariance matrix can be written as 
\begin{eqnarray}
    C_{ij} = \begin{pmatrix}
                  \sig_1^2 & 0 & 0 & \cdots & 0\\
                  0 & \sig_2^2 & 0 & \cdots & 0 \\
                  \vdots &  & \ddots &  & \vdots \\
                  0 & \cdots &  & C_{ij}^{\rm WiggleZ} \cdots& 0  \\
                  \vdots & \cdots & &\cdots& \vdots \\
                  0 & \cdots &  & \cdots & 0
             \end{pmatrix} \, ,
\end{eqnarray}
where, $C_{ij}^{\rm WiggleZ}$ is the covariance matrix of the WiggleZ measurements which is given by
\begin{eqnarray}
    C_{ij}^{\rm WiggleZ} = 10^{-3} \begin{pmatrix}
    6.400~ & 2.570~ & 0.000 \\
    2.570 & 3.969 & 2.540 \\
    0.000 & 2.540 & 5.184
    \end{pmatrix} \, .
\end{eqnarray}

\subsection{Results}

\begin{table}[h]
\begin{center}
\caption{Observational constraints of the parameters for $\Lambda$CDM, $w$CDM, CPL, $Pf1$, $Pf1+Phantom$ and $Pf2$ models are given along with the corresponding model comparison statistics with AIC and BIC. We have usd Eqs.~\eqref{eq:w0_sc} and \eqref{eq:w0} to constrain the value of $w_0$ for the parametrizations $Pf1$, $Pf1+Phantom$ and $Pf2$.}
\label{tab:cons}
\resizebox{\columnwidth}{!}{%
\begin{tabular}{c c c c c c c}\hline \hline
 Param. &  $\Lambda$CDM  &  $w$CDM &  CPL &  $Pf1$  &  $Pf1+Phantom$ & P$f$2\\ 
\hline \hline
 $\Om_{\m 0}$ & $0.3129\pm0.0067$ & $0.3169^{+0.0056}_{-0.010}$ & $0.3124\pm 0.0077$ & $0.3168\pm 0.0073$ & $0.3148\pm 0.0076$ & $0.3168\pm 0.0073$  \\ 
$h$ & $0.6757\pm 0.0051$ & $0.6734\pm 0.0073$ & $0.6774 \pm 0.0076$ & $0.6704^{+0.0064}_{-0.0058}$  & $0.6733\pm 0.0072$ & $0.6705\pm 0.0062$ \\
 $\omega_{\rm b}$ & $0.02240\pm 0.00014$ & $0.022507^{+0.000068}_{-0.00023}$ & $0.02236 \pm 0.00015$ & $0.02245\pm 0.00014$ & $0.2242\pm 0.00014$ & $0.02245\pm 0.00014$   \\
 $r_{\rm d} h$ & $100.46\pm 0.70$ & $100.1\pm 1.2$ & $100.2 \pm 1.1$ & $99.68^{+0.92}_{-0.80}$ & $100.1\pm 1.0$ & $99.66^{+0.90}_{-0.81}$   \\
 $\sig_8$ & $0.746\pm 0.029$ & $0.744\pm 0.030$ & $0.751 \pm 0.030$ & $0.744\pm 0.029$ & $0.745\pm 0.029$ & $0.744\pm 0.029$    \\
 $w_0$ & $---$ & $-0.974^{+0.014}_{-0.042}$ & $-0.873 \pm 0.067$ & $<-0.96$  & $-0.988\pm0.028$ & $-1.008^{+0.059}_{-0.110}$ \\
 $w_{\rm a}$ & $---$ & $---$ & $-0.55^{+0.31}_{-0.28}$ & $---$  & $---$ & $---$ \\
 $\Om_{\delta}$ & $---$ & & $---$ & unconstrained & $>-0.3$ & $>1.37$     \\
$\Om_{01}$ & $---$ & $---$ & $---$ & $---$ & $---$ & unconstrained \\
$\al_2$ & $---$ & $---$ & $---$ & $---$ & $---$ & $0.082^{+0.028}_{-0.076}$ \\
$\al_1$ & $---$ & $---$ & $---$ & $<0.1$ & $0.033\pm 0.083$ & $>0$ \\
$M$ & $-19.434\pm 0.015$ & $-19.434^{+0.014}_{-0.024}$ & $
19.416 \pm 0.022$ & $-19.445 \pm 0.017$  & $-19.439\pm 0.018$ & $-19.439\pm 0.015$ \\
\hline 
$\chi^2_{\rm min}$ & $1447.00$ & $1448.18$ & $1443.72$ & $1447.11$ & $1446.82$ & $1447.04$  \\
$\chi^2_{\rm red}$ & $0.90$ & $0.88$ & $0.88$ & $0.88$ & $0.88$ & $0.88$  \\
AIC & $1459.01$ & $1462.18$ & $1459.72$ & $1467.04$ & $1462.82$ & $1466.88$  \\
$\Delta$AIC & $0$ & $3.17$ & $0.71$ & $4.10$ & $3.81$ & $8.00$  \\
BIC & $1491.40$ & $1500.05$ & $1503.00$ & $1506.39$ & $1506.10$ & $1521.14$  \\
$\Delta$BIC & $0$ & $8.65$ & $11.60$ & $14.99$ & $14.70$ & $29.74$  \\ 
\hline \hline
\end{tabular}}
\end{center}
\end{table}

\begin{figure}[ht]
\centering
\includegraphics[scale=0.8]{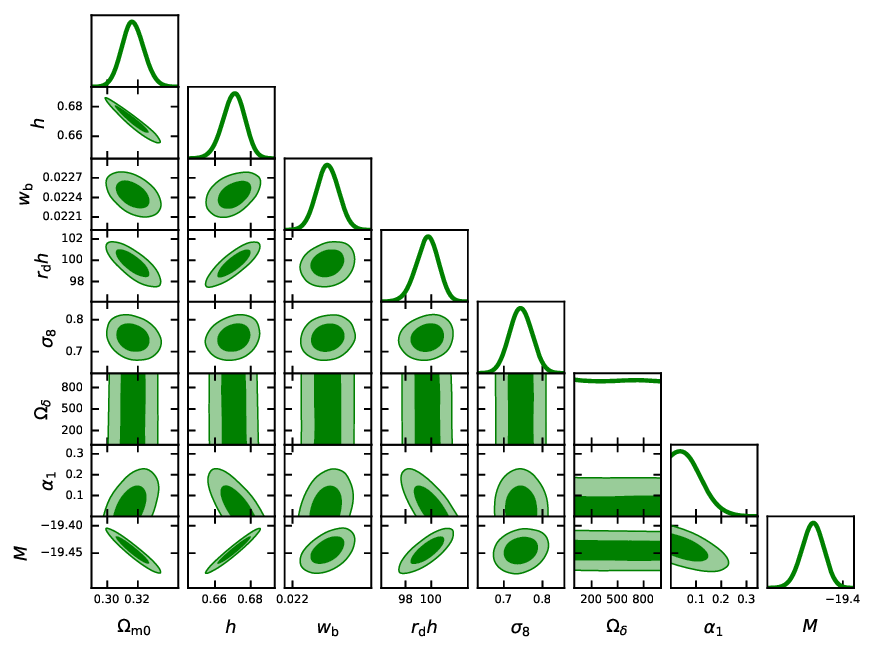}
\caption{1$\sig$ and 2$\sig$ confidence levels of the model parameters for the $Pf1$ model.}
\label{fig:Pf1}
\end{figure}

\begin{figure}[ht]
\centering
\includegraphics[scale=0.8]{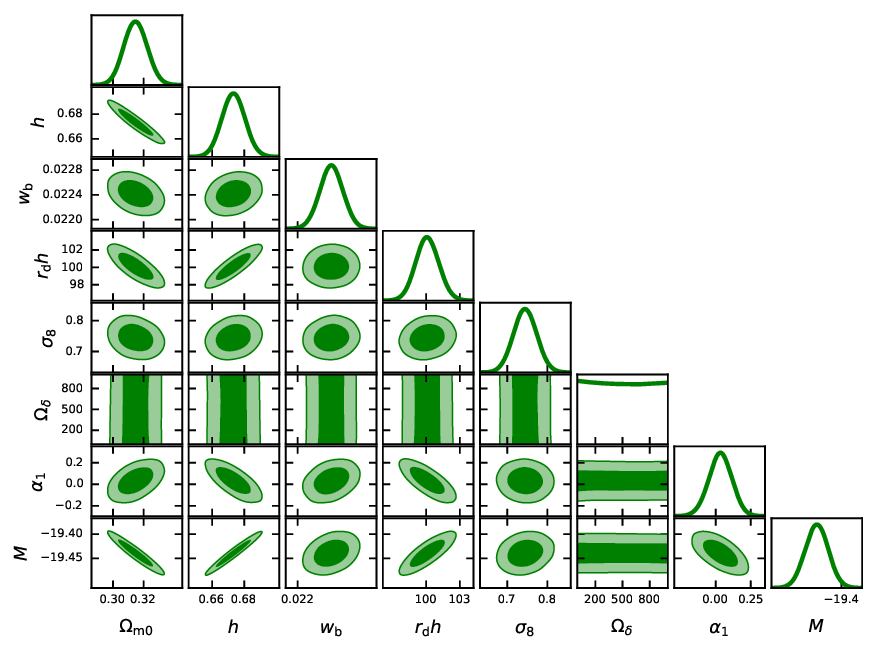}
\caption{1$\sig$ and 2$\sig$ confidence levels of the model parameters for the $Pf1+Phantom$ model.}
\label{fig:Pf1phant}
\end{figure}

\begin{figure}[ht]
\centering
\includegraphics[scale=0.8]{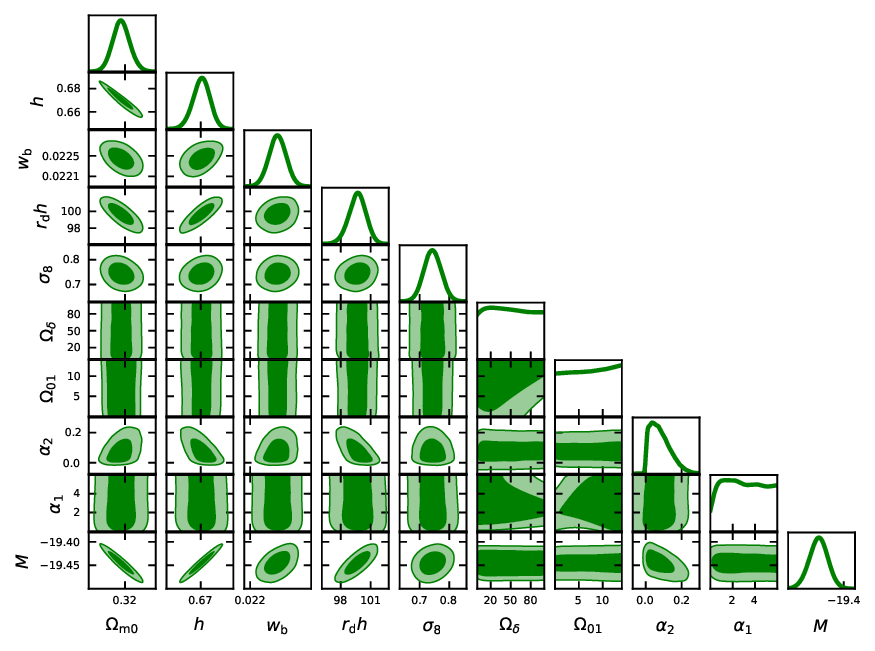}
\caption{1$\sig$ and 2$\sig$ confidence levels of the model parameters for the $Pf2$ model.}
\label{fig:Pf2}
\end{figure}

We present the results of the cosmological data analysis for the data combination ${\rm CMB}+{\rm BAO}+{\rm PP}+{\rm Hubble}+{\rm RSD}$ in Tab.~\ref{tab:cons}. The contours of the parameters for the models $Pf1$, $Pf1+Phantom$ and $Pf2$ are shown in the Figs.~\ref{fig:Pf1}, \ref{fig:Pf1phant} and \ref{fig:Pf2} respectively. From the values of $\Delta {\rm AIC}$ and $\Delta {\rm BIC}$ we can say that the considered data prefers the standard $\Lambda$CDM model over the other five models. Among the five parametrizations considered for the data analysis CPL appears to be the more favoured by the data. $Pf1+Phantom$ and $w$CDM parametrizations are equally favoured after the CPL parametrization. Among all the models the less preferred model is $Pf2$. For $Pf1+Phantom$ we consider the prior of $\al_1$ as $\{-2,2\}$ to incorporate the phantom region. If we compare this model with the $Pf1$ model then it is very clear that the model $Pf1+Phantom$ is more preferred by the considered data compared to $Pf1$ as the values of $\Delta {\rm AIC}$ and $\Delta {\rm BIC}$ are less for the $Pf1+Phantom$ model. So, when we restrict the scenario only within the non-phantom region the data seem to disfavour the model. Inclusion of phantom region makes the scenario more preferable even though the standard $\Lambda$CDM model still remains as the best model. Here we should mention that the parametrization~\eqref{eq:rho_para} is mainly for quintessence field. So, for $f=1$ case, considering the value of $\al_1<0$ can lead to phantom behaviour but proper investigation should be done while using the parametrization~\eqref{eq:rho_para}. So basically if we consider thawing kind of dynamics then, we see, that the data prefers to have phantom region within $1\sigma$ bound. This has to be noted that even after considering phantom region for $Pf1$ case, $\al_1=0$, which represents a constant energy density or CC, is well within the $1\sig$ bound. This gives an interesting result as from Tab.~\ref{tab:cons} we can see that the CC is not preferred only for CPL parametrization. In all other parametrizations CC is well within the $1\sig$ bound. So, the evidence of the dynamical dark energy \cite{DESI:2024mwx,DESI:2024aqx,DESI:2024kob} over CC is present only in the CPL parametrization. In other words, the evidence of dynamical dark energy over $\Lambda$CDM is parametrization dependent.  For the models $Pf1$ and $Pf1+Phantom$, from Figs.~\ref{fig:Pf1} and \ref{fig:Pf1phant}, we can see that the parameter $\Om_\delta$ is not well constraint except having a lower bound around $-0.3$ for $Pf1+Phantom$ model. This implies that we can not precisely talk about the values of $z_1$ (Eq.~\eqref{eq:z1_thaw}). For the parametrization with $f=2$, {\it i.e.}, $Pf2$ we see that the parameter $\Om_{01}$ is unconstrained and the parameters $\Om_\delta$ and $\al_1$ have bounds but the data can't fully constrain them (Fig.~\ref{fig:Pf2}). We have calculated the the bound on the present value of dark energy EoS $w_0$ from the Eq.~\eqref{eq:w0_sc} by using the bounds on the model parameters. So similar to the case to $Pf1$ for $Pf2$ also we can not precisely talk about the values of $z_2$ (Eq.~\eqref{eq:z2_sc}) and $z_1$ (Eq.~\eqref{eq:z1_sc}). We hope in future we will be able to constrain these parameters with more precise data. We have not considered the phantom region for $Pf2$ as that has to be investigated properly first for the higher redshifts.

\section{Discussions and Conclusions}
\label{sec:conc}
In this paper, we present a general parametrization~\eqref{eq:rho_para} for the energy density of the quintessence field. The parametrization~\eqref{eq:rho_para} can successfully mimic all kind of scalar field dynamics, namely scaling-freezing, tracker and thawing dynamics which has been shown in the Secs.~\ref{sec:scaling-freezing}, \ref{sec:tracker} and \ref{sec:thawing} respectively. It can also mimic the dynamics for an oscillatory potential either by choosing very large value of $f$ or by considering the average EoS of the scalar field.  So, our parametrization can mimic the dynamics of a quintessence field for a particular potential as well as it can also behave as model-agnostic approach to work with scalar field models. This parametrization can also work as a parametrization for phantom scalar fields if we do not consider much higher redshifts. For higher redshifts it has to be investigated properly for the phantom cases. Now, while it is interesting that the parametrization~\eqref{eq:rho_para} can mimic any kind of quintessence dynamics it also needs more parameters to represent more complicated dynamics like scaling-freezing or tracker. Having more free parameters makes the scenario less interesting from the cosmological data analysis point of view as the current data may not be able to constrain all the parameters properly which can be seen in Fig.~\ref{fig:Pf2}. But we expect that in near future we will have much more precise data and we will be able to constrain more free parameters. If we can do that then our parametrization will properly point towards the actual dynamics. Apart from this issue the main advantage of our parametrization is that it can mimic the actual dynamics of a quintessence field for a particular potential and it reduces the computing time for data analysis to a very significant amount. When we work with scalar fields the computation for data analysis is, in general, very time consuming and as long as our computation is concerned it takes around $70s/iteration$ after including the flatness condition which makes the code slower. In this regard, the parametrization~\eqref{eq:rho_para} can make it much more affordable computationally and for our code it takes around $10s/iteration$ which is similar to the $\Lambda$CDM case. We should also mention here that the performance of the code depends on the optimisation. So, the comparison is completely based on our codes.

We have compared our scenario with the standard $\Lambda$CDM, $w$CDM and CPL models using the recent cosmological data. In this regard we have considered $f=1$ and $f=2$. For $f=1$ case we have considered both non-phantom ($Pf1$) and phantom ($Pf1+Phantom$) regions. For $f=2$ ($Pf2$) we have considered only non-phantom regions. Our results tell us that the cosmological data prefers $\Lambda$CDM model over other models. In fact, even though we have considered the DESI 2024 DR1 data of BAO the CC is well within the $1\sig$ bound. In fact, only CPL parametrization shows a preference of dynamical dark energy over $\Lambda$CDM. $\Lambda$CDM consistent with all other parametrization considered in Tab.~\ref{tab:cons}. Similar results have also been achieved recently in \cite{Sakr:2025daj}. Interestingly, the scenario $Pf1$, which is non-phantom consideration with thawing dynamics, is less preferred over the $Pf1+Phantom$ which is with the phantom region. This result is consistent with the results obtained in \cite{Vagnozzi:2019ezj,Vagnozzi:2018jhn,Teng:2021cvy,Ramadan:2024kmn} but contradicts the findings of \cite{DESI:2024kob}.

\begin{acknowledgments}
MWH acknowledges the High Performance Computing facility Pegasus at IUCAA, Pune, India, for providing computing facilities. The authors acknowledge the financial support from ANRF, SERB, Govt of India under the Start-up Research Grant (SRG), file no: SRG/2022/002234.
\end{acknowledgments}

\bibliographystyle{JHEP}
\bibliography{references}

\end{document}